\documentclass[conference]{article}

\usepackage{arxiv}

\usepackage[utf8]{inputenc}
\usepackage[T1]{fontenc}    

\usepackage{amsmath}
\usepackage{graphicx}
\usepackage{siunitx}
\usepackage{hyperref}
\usepackage{fancyref}
\usepackage{todonotes}
\newcommand{\khz}[1]{\SI{#1}{\kilo\hertz}}
\newcommand{\khzr}[2]{\SIrange{#1}{#2}{\kilo\hertz}}


\usepackage{stfloats}

\begin{document}
%
\title{BatNet: Data transmission between smartphones over ultrasound}

\author{
Almos Zarandy\\
Computer Laboratory\\
University of Cambridge\\
\And
Ilia Shumailov\\
Computer Laboratory\\
University of Cambridge\\
\And
Ross Anderson\\
Computer Laboratory\\
University of Cambridge\\
}

\maketitle

\begin{abstract}
In this paper, we present BatNet, a data transmission mechanism using ultrasound signals over the built-in speakers and microphones of smartphones. Using phase shift keying with an 8-point constellation and frequencies between \SIrange{20}{24}{\kilo\hertz}, it can transmit data at over \SI{600}{\bit\per\second} up to \SI{6}{\metre}. The target application is a censorship-resistant mesh network. We also evaluated it for Covid contact tracing but concluded that in this application ultrasonic communications do not appear to offer enough advantage over Bluetooth Low Energy to be worth further development. 
\end{abstract}

\section{Introduction}
\label{sec1}
\noindent An important aspect of networks is resilience -- the ability to maintain service in the face of accidental or deliberate disruption. The ubiquity of mobile phones has made people dependent on wireless networks, whether Wi-Fi or cellular. Radio networks can be jammed by technical means, or a service can be switched off --  during a protest, the authorities may switch off cellular networks and in extreme cases may even cut off the wireline Internet to disable Wi-Fi too. Power cuts can also disable wireless networks. Whether a service failure is caused by political protest, a natural disaster or the failure of an electricity substation, people have to improvise alternative means of communication. 

\noindent We explore the use of ultrasound as a fallback communication channel. It can be jammed just like radio, but doing so in a large area requires large loudspeakers; it also causes a health hazard as intense ultrasound may cause hearing loss or even death. The authorities in most countries might be more reluctant to do this than they are to interfere with wireless networks. We therefore explore whether smartphones can be used for ultrasound communication. 

\noindent The spectrum of human hearing usually goes from \SI{20}{\hertz} up to about \khzr{18}{20} for young people, with older people progressively less able to hear higher frequencies.  Most smartphones have a sampling frequency of \khz{48}. We find that some frequencies in the near-ultrasound range of \khzr{20}{24}
can often be used for messaging.  However both playback and recording quality differ by frequency, and by device, and are generally weaker than in the audible range. 

\noindent The wavelength of sound in this region is about \SI{1.7}{\centi\metre}, so even small movements of a device may introduce a significant Doppler shift that causes a temporary change in frequency and a permanent shift in phase. This is a significant operational limitation because phones are held in hands or pockets and are never absolutely still. The application can thus only transmit short messages that can be finished before the devices move significantly, and it must cope with shifts in the carrier phase. 

In this paper we made the following contributions
\begin{itemize}
    \item We have implemented a signal generator that modulates data at over 600bps on the ultrasonic spectrum beyond the range of human hearing. 
    \item We have implemented a matching signal processor that recognises and decodes this signal. 
    \item We evaluated the performance of the resulting communications channel. We found significant angular dependency: the signal can carry up to \SIrange{6}{8}{\metre} in the direction the speaker is facing, but possibly less than \SI{1}{\metre} in other directions. 
    \item We evaluated the possible use of ultrasonic communications in covid contact tracing, but concluded that there is not enough advantage over BLE to justify further development. 
    \item We have released our code as an Android demo app.\footnote{\url{https://github.com/zarandya/batnet}}
\end{itemize} 

\section{Related work}

\noindent There have been attempts to use ultrasound for communication, initially in water with submarines, then in air using specialised microphones. The first attempts with smartphone microphones were beaconing products.   {\it Shopkick}\footnote{\url{https://www.shopkick.com}} has transmitters inside shops that emit ultrasound and Bluetooth Low Energy beacons; an app detects them to learn that the phone user entered the shop. {\it Silverpush}\footnote{\url{https://www.silverpush.co}} detects ultrasonic beacons hidden in TV adverts to monitor user habits. Both of these products raised privacy concerns~\cite{7961950}.

\noindent {\it Lisnr}\footnote{\url{https://lisnr.com}} is a company that provides ultrasonic data transmission. Their proprietary technology has been deployed in mobile payment and ticketing systems, but they do not provide a free API. 

\noindent {\it Google Nearby}\footnote{\url{https://developers.google.com/nearby/}} is a technology that allows users to discover nearby devices and send messages to them. It uses a combination of Bluetooth, Wi-Fi, and ultrasound, and does not work without the radio channels. \\

%


\section{Implementation}

\begin{figure*}[tbh]
	\centerline{\includegraphics[scale=1.6]{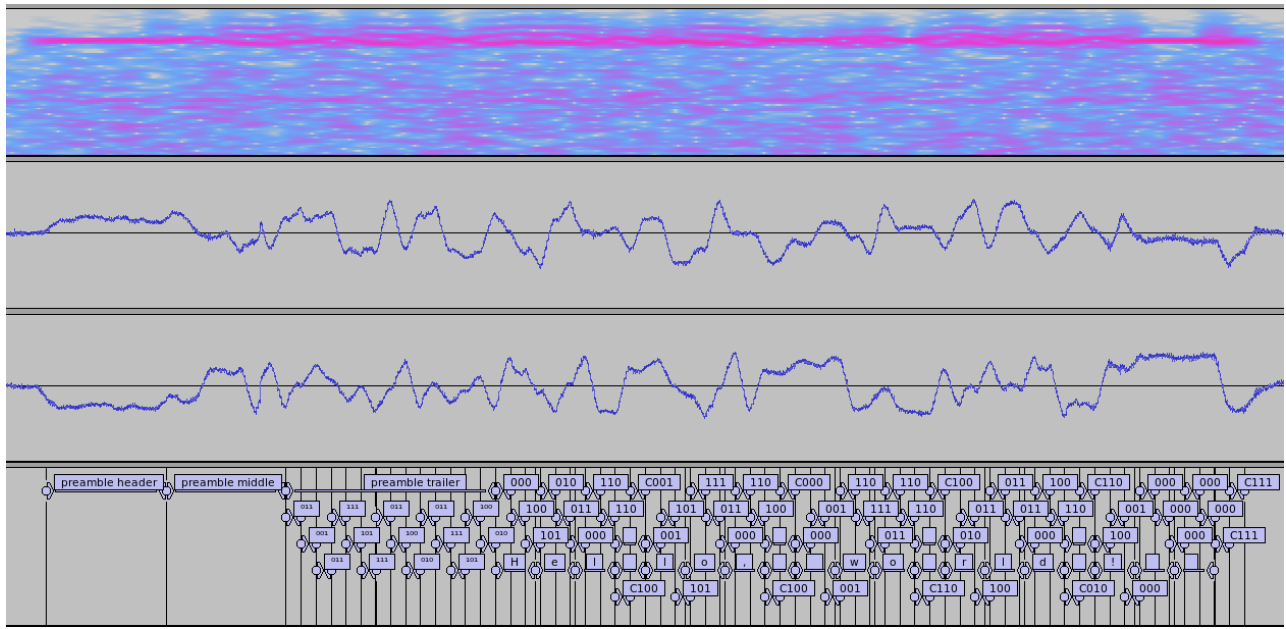}}
	\caption{Example signal containing payload \c{"Hello, world!"} sent at frequency \khz{22.5}. The	top row is the spectrogram of the recorded sound signal, the middle two rows are the real and imaginary parts of the demodulated signal.}
\label{fig:helloworld}
\end{figure*}

\subsection{Modulation scheme}

\noindent The modulation scheme used by BatNet is {\em phase-shift keying} (PSK), which carries data by changing the phase of a constant frequency signal. Eight distinct phases can be differentiated with good accuracy, allowing a symbol for every three bits~\cite{Proakis}. To demodulate PSK, the (real) signal is multiplied with the reference (complex) signal. In the Fourier domain, this shifts the symbol at frequency $-f_c$ to zero, and at frequency $f_c$ to $2f_c$. Then the higher frequency copy is filtered out, and the symbol is represented by the DC term. 

\noindent Amplitude modulation would not have been a good choice because having lower amplitude symbols in the signal would further reduce the already short range. Frequency modulation would have worked, but it could be problematic due to different phones responding better or worse to different frequencies in the ultrasonic spectrum. Google Nearby uses frequency spreading to overcome this, but that results in inefficient use of spectrum as redundant data is sent over multiple frequencies. 

\noindent PSK has its own problems. The signal is discontinuous at the symbol boundary,  which introduces softly audible clicks, and the speakers are slower to adapt to a change in phase then to a change in frequency, which introduces additional noise at the symbol boundaries and further limits minimum symbol size. Furthermore, a Doppler shift changes the carrier phase, resulting in symbols being recognised incorrectly. 

\subsection{Signal Format}

\noindent The signal consists of a discrete sequence of symbols, where the phase is constant. There are eight possible symbols which have constant length and there is a constant length transition time between them. To avoid discontinuities in the signal between the symbols, a different frequency is used as a transition between symbols where the phase will change. This transition frequency is ignored by the receiver, and only the symbols are sampled. 

\noindent The signal starts with a preamble to synchronise carrier phase and symbol boundary, and to inform the receiver(s) that a signal is coming. The first part of the preamble has a constant reference frequency and carrier phase. Then there is a {\it synchronisation sequence}, and a trailer that is used to discard false-positive preambles. After the preamble, the symbol boundary is fixed, but the carrier phase can still be refined using error correction. 

\noindent Then comes the data. Each symbol corresponds to three bits of data, arranged in an eight-point constellation with Grey coding, which means that adjacent symbols differ only in one bit. Some of the data bits are redundant and used for error correction. The scheme can work with multiple rates of redundant bits. 

\begin{figure*}
	\centerline{%
		\scalebox{.7}{%
			\input{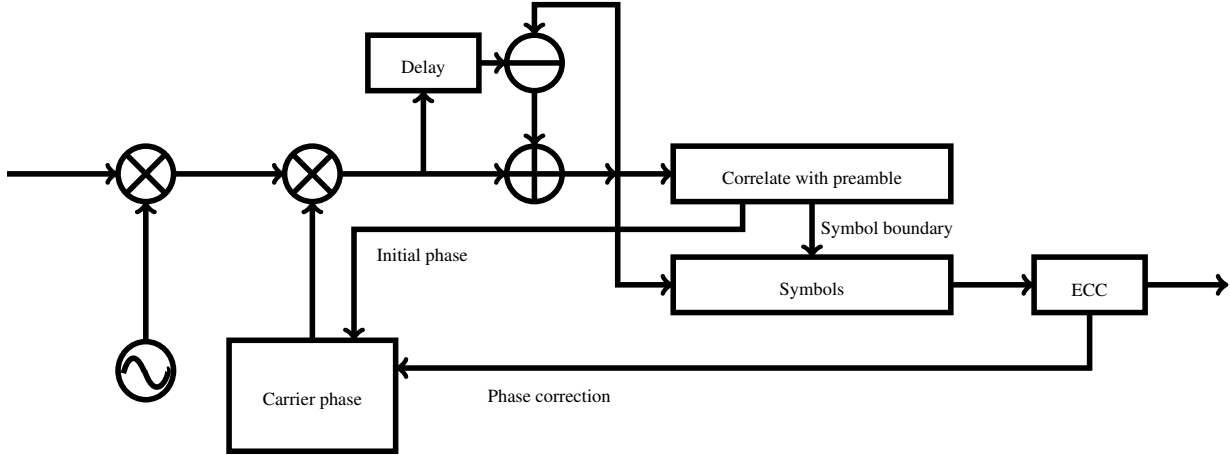}
		}
	}
	\caption{Receiver implementation}
\end{figure*}

\subsection{Demodulation}

\noindent The signal is demodulated by multiplication with the reference symbol at the prevalent near-ultrasound frequency, and windowed averaging removes the high-frequency component. When the symbol boundary is known, the signal is split there. Otherwise the signal is searched for the {\it synchronisation sequence}, and if it is found, the symbol boundary and reference phase are determined. 

\subsection{Error correction}

\noindent Redundant bits are added to the data to correct errors. We use a {\em cyclic redundancy check} (CRC)~\cite{mcdaniel_mcdaniel}. 

\noindent The most common cause of errors other than noise is a skew in the carrier phase between the two devices. This can be introduced by a clock drift between the phones, or (most often) by a change in their distance during signal transmission. This means that phase errors in a sequence of symbols are not independent, and error correction has to be done on the demodulated signal rather than the symbols. 

\noindent The error correction returns not just the received data but also a suggested correction to the carrier phase. This suggestion is applied before the next iteration, so drifts in the carrier phase can be corrected. Otherwise symbols would get replaced by the next symbol in the constellation. 

\subsection{Calibration}

\noindent BatNet supports multiple frequencies in the near-ultrasound range. The transmission quality depends on the characteristics of the speaker of the sender, the microphone of the receiver, and it might also depend on context. Calibration is required to find the ideal frequency between every pair of devices. \\

\section{Evaluation}

\noindent Evaluation was performed on the four devices listed in~\Fref{tab:phonespec}. Other devices were involved in initial tests, but these were the main devices used for evaluation. 

\begin{table*}
\centering
\caption{Specification of phones used for evaluation.}
\label{tab:phonespec}
\begin{tabular}{|ll|ll|lll|}
	\hline
	Manufacturer&Name&Android version&Year&Speaker rate&Microphone rate&\\\hline
	Samsung&Ace 3&4.2.2 Jelly Bean&2013&\khz{44.1}&\khz{48}&\\
	LG/Google&Nexus 4&5.1.1 Lollipop&2012&\khz{48}&\khz{48}&\\
	LG/Google&Nexus 5&6.0.1 Marshmallow&2013&\khz{48}&\khz{48}&\\
	Nokia&5.1&9 Pie&2018&\khz{48}&\khz{48}&\\
	\hline
\end{tabular}
\end{table*}

\noindent In most tests, one of these devices was the transmitter while the other three were the receiver.  Unless otherwise specified, the devices were placed on cardboard boxes on top of a soft surface, and the speaker of the sender was set to face the microphone of the receiver. 

\noindent All evaluation was performed outdoors. There was some noise in the environment (cars, birds, dogs, grass cutting, welding, chainsaw etc.) but these don't produce sound in the ultrasonic region so they didn't jam the signals. The only thing that seemed to have a negative effect was small aeroplanes flying low, which aren't common except under a general flying area or in the vicinity of an airport\footnote{We leave it to future work to check whether police helicopters could degrade Batnet; we did not have access to a helicopter for testing.}.

\noindent The receivers in the test wrote the raw sound data to a file, which was then evaluated on a computer. The signal processor can run on the phones in real time.

\noindent The test results are presented as {\it transmission quality}, which is for our purposes the proportion of correctly received symbols. More precisely, for each symbol, the closest constellation point is found. If it is the correct one, it scores one point, while if it is an adjacent one, it scores half a point. The average of these scores is the transmission quality. 

\subsection{Performance and Power usage}

\noindent Running BatNet increases battery use, as it performs a significant continuous computational workload and emits sound signals. On the Nexus 5, it will drain $40\%$ of the battery in three hours of active usage. That is about \SI{1}{\watt}, similar to classic Bluetooth. In comparison, Bluetooth Low Energy (BLE) uses \SIrange{0.01}{0.5}{\watt} depending on the use case.  For  a mesh messaging network used in a protest, this will be usable if the user fully charges the phone before. For emergency communications following a natural disaster, we expect that mobile phones can be recharged using cars.

\noindent Battery life can be improved by putting the signal reading thread to sleep when it is not in use. For example, the phone could wake up and listen for six seconds every minute at random. This would introduce latency as two phones could contact each other only when both were active, but it could be reduced if a phone anxious to communicate were to transmit continuously.  Alternatively the signal-reading thread could be put to sleep when we know messages cannot be successfully received -- as when there's a lot of noise or the phone is moving rapidly. We leave such optimisations for future work.

\noindent The signal processor runs in real time, but to achieve this it had to be written in native C code. The computationally intensive part is error-correcting code resolution, but by choosing a small ECC block size it becomes manageable. 

\subsection{Transmission range}

\begin{figure}
    \centering
	\scalebox{0.5}{\input{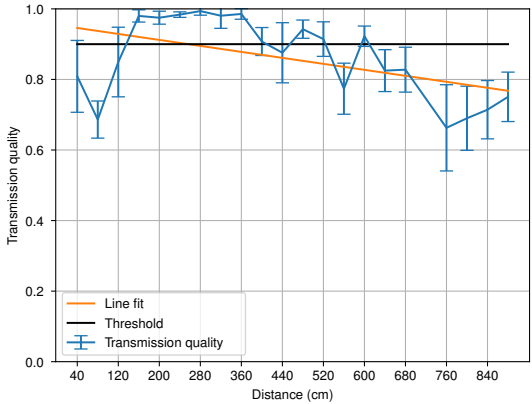}}
	\caption{Dependence of transmission quality on data from the Nokia 5.1 device recorded by the Nexus 4 with changing distance. Transmission quality degrades linearly with distance; see the fit orange line. It still can produce acceptable quality from up to \SIrange{6}{8}{\metre}}
	\label{fig:changedist}
\end{figure}

\noindent To measure transmission range, the transmitter was placed at a fixed distance from receivers with the speaker facing the receiving microphones. A signal was sent 16 times, then the distance was increased. The same frequency was used throughout the test -- the one that scored highest in calibration for those devices. Carrier-phase drift correction was turned off. 

\begin{figure}
    \centering
	\scalebox{0.5}{\input{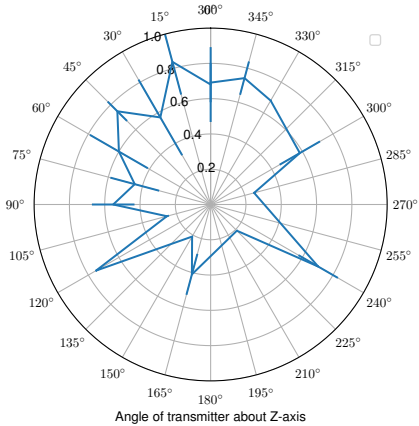}}
	\caption{Dependence of transmission quality on angle of transmitter to the direction of the other device. Transmission quality is mostly dependent on angle of the transmitter and is best in the direction of the speaker. It is also dependent on the direction of the microphone but not as much as that of the speaker.}
	\label{fig:changeangle}
\end{figure}

\noindent The signal carries well to about \SIrange68\metre in the direction the speaker is facing, as can be seen on~\Fref{fig:changedist}. In the opposite direction, the signal does not carry that well (see~\Fref{fig:changeangle}). 

\subsection{Symbol rate}

\begin{figure}
	\centering
	\scalebox{0.5}{\input{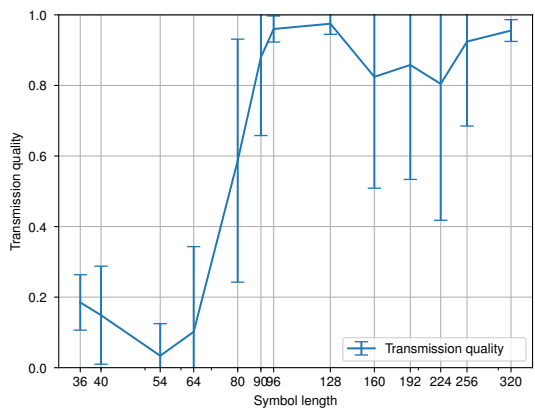}}
\caption{Transmission quality with different symbol lengths. }
\label{fig:changelen}
\end{figure}

\noindent Due to the low sampling frequency, a symbol has to be many times longer than the wavelength. There is also a transition phase between symbols. To determine the best symbol size, a series of experiments was run with varying symbol length, carrier phase drift correction being once more disabled. As seen on~\Fref{fig:changelen}, for still devices, the transmission quality performs consistently well at symbol lengths above a threshold of 96 samples, but breaks down under this. 

\noindent For hand-held devices which move relative to each other, a longer symbol of 160 samples is advisable. This will result in a raw throughput of \SI{900}{\bit\per\second}, but some of these bits are then used for error correction. 

\subsection{Error correction}

\noindent In the above sections, transmission quality was evaluated without carrier phase adjustment.  In order for error correction to work, the transmission quality has to be above $0.9$.

\noindent We experimented with multiple low-degree CRC polynomials (the degree of the polynomial is the number of redundant bits added to each block). Error correction has to be performed on small blocks, because algorithms for finding the optimal carrier phase adjustment are slow. So only a small number of redundant bits can be added to each block, and a low-degree polynomial is required. We tried polynomials of degree 3, 4, 5, and 6, and higher-degree polynomials generally achieved better accuracy. The final version uses a degree-5 polynomial with 16 data bits and 5 checksum bits. 

\subsection{Censorship circumvention}

\noindent The application is susceptible to jamming, but unless the opponent uses high volume, the jamming range is short.  We tried sending messages between two phones while using a third phone to jam the carrier frequency. We could jam receivers up to about \SI{3}{\metre} in the direction of the speaker of the jammer or \SI{1.5}{\metre} in other directions, if jamming was successful, the message wasn't received, otherwise it didn't seem to affect transmission quality. In a protest, an adversary can temporarily jam small regions, but to jam a significant area, the authorities would need to infiltrate a large number of adversaries, or subvert a lot of people's phones, or perhaps use a low-flying aircraft. 

\noindent BatNet can operate on multiple frequencies, so if one frequency is jammed, another frequency can be used. Jamming all Batnet frequencies might involve barrage jamming, which would produce noise in all near-ultrasound frequencies, with attendant health risks near the loudspeakers; another possibility might be a squidging oscillator, which would rapidly cycle through the frequencies BatNet can use. Bluetooth uses frequency hop to provide some jam-resistance; such techniques could be used here as well, though the number of usable frequencies is much more limited than in radio. 

\subsection{Comparison with similar products}

\noindent As already noted, there are two products that implement device-to-device ultrasound message passing similar to BatNet: Google Nearby and Lisnr.~\Fref{tab:comparison} compares their performance.

\begin{table*}
	\centering
\begin{tabular}{|l|r|r|l|}
	\hline
	&bandwidth&Maximum distance&modulation scheme\\\hline
	Google Nearby&\SI{94.5}{\bit\per\second}&\SI{10}{\metre}&DSSS\\\hline
	Lisnr (KAB)&\SI{1000}{\bit\per\second}&\SI{1}{\metre}&FSK\\\hline
	BatNet&\SI{685.7}{\bit\per\second}&\SIrange{6}{8}{\metre}&PSK\\\hline
\end{tabular}
\caption{Comparison of data-over-audio products}
\label{tab:comparison}
\end{table*}

\noindent The ultrasonic communication used by Google Nearby has a throughput of \SI{94.5}{\bit\per\second}~\cite{8080245}. It uses {\em direct-sequence spread spectrum} (DSSS) which has better resilience to Doppler shifts and might therefore be more reliable on moving devices. The cost of this is that the DSSS chip rate must be higher than the data rate, limiting the data rates that can be achieved.  It is not clear what parts of Google Nearby use ultrasound, but it does not work without Bluetooth and Internet connections. 

\noindent Lisnr claims a throughput of \SI{1}{\kilo\bit\per\second} with their recently launched KAB profile, which works up to 3 feet between mobile devices. Other profiles can transmit to higher distances but at lower rates. This is used to validate a ticket in one second or to process an EMV payment in about 1.5 seconds. It uses frequency modulation, according to Daniel Arp and colleagues who reverse engineered it~\cite{7961950}. This suggests that transmission quality may not depend on frequency as much as we initially feared. 

\subsection{Testing with handheld devices}

\noindent The testing reported in the previous sections was performed with static devices. In practice, devices would be held in a hand or a pocket. Hands shake and humans move; the phone moves with them, and messages sent between two phones that are moving too quickly are lost. A possible improvement would be to use the phone's sensors to detect when it is still, but this would need care; it could deny service needlessly between two phones moving together (e.g. in the same bus).

\section{Ultrasound communication for COVID Contact Tracing}

\noindent While BatNet was developed with a censorship-resistant mesh network similar to FireChat in mind, the coronavirus pandemic caused a surge of interest in public-health technologies. Apps have been developed for public information, symptom tracking, quarantine enforcement, contact tracing and combinations of these. Most fielded contact tracing apps use BLE beacons for device discovery, but this has run into significant problems because of the restrictions that Apple in particular have imposed. In order to protect user privacy, apps can only use BLE if they are in foreground. This led, for example, to the abandonment of the UK contact-tracing app after it emerged that only 4\% of iPhone users were detected~\cite{Lovett2020}.

\noindent Might ultrasound be used as an alternative? 
To answer this question, we investigated the limitations of BatNet in this context. 

\begin{itemize}
    \item It has a short range, which depends critically on the orientation of the transmitter and receiver. 
    \item It can only be used to pass short messages, because of carrier phase drift. 
    \item Message passing is not completely reliable, particularly in noisy environments such as buses, trains and pubs. 
    \item We didn't test with iPhones but we understand that only one app can use the Apple screen/audio capture API at once, so as with BLE it could be flaky on their devices, needing user interaction.
    \item The application drains a typical phone battery in about 7h if operated continuously.
\end{itemize}

\noindent The reliability issue can probably be overcome by repeating messages. If two people are close together for the 15 minutes considered by most app operators to count as an epidemiologically significant contact, their devices will send enough beacons that some of them will be received. There are likely to be issues in noisy environments, such as tube trains with screeching wheel noise, leading to missed alarms; there might even be false alarms over voice messaging systems that did not limit transmit bandwidth to 20KHz. The battery life issue can also be mitigated, by using a beaconing protocol with a duty cycle of 10\% or less.

\noindent The variable range is an issue, just as with BLE. If two people hold their phones facing each other, the range can be as much as 10m, which will give rise to false alarms; if they both have their phones in pockets facing their bodies, detection may fail completely, causing missed alarms. BLE has similar problems, in that the range also depends on how phones are orientated with respect to each other and their users' bodies: if the received signal strength indicator is set to give reliable detection at 2m then it will often detect other phones at 10m, leading to false alarms. 

\noindent In both cases, there will be further missed alarms if both people are not simultaneously running the contact-tracing app, whether out of privacy concerns, because of the need for manual intervention on iPhones, to prolong battery life, or because of the perverse incentives. Experience over many countries suggests that most people won't run such an app, so most alarms are missed. In any case, contact tracing appears to work best when it's both personal and local~\cite{Bounds2020}. We do not propose to summarise the voluminous literature on contact tracing at any length here. One of us wrote an early critique that predicted some of the difficulties~\cite{Anderson-contact2020}. Of course, in a crisis it's reasonable to try everything, but it makes sense to fail fast -- to abandon lines of inquiry that turn out to be unhelpful.

\noindent Our assessment was and remains that, in most circumstances, ultrasonic communications offer contact tracers no advantage over BLE, and in the circumstances of most interest -- locating people who were sitting close enough to the index case for long enough but who are unknown to them, such as nearby strangers in a bus, train or pub -- the level of noise may make the performance worse. There are some rare circumstances where ultrasonic tracing works better, such as on older devices. It might also be possible to measure distance better than with BLE by using time-of-flight; we did not investigate this because of the complexity. But our overall assessment was that ultrasonic contact tracing is unlikely to be useful, except perhaps in a complex multi-channel tracing app as a secondary or fallback mechanism. Having reported this to our contact on the UK government team, we discontinued this line of research.

\noindent Although we considered BatNet as a means of communication during emergencies, when normal networks are disrupted because of power outages, the range we obtain using commodity smartphones is so short that we cannot see any advantage over using short-range radio and building mesh networks over that.

\noindent BatNet may well, however, be viable as a communications channel in niche applications such as in urban protest. There, participants may be motivated to run a special app, unlike in contact tracing where many people have got fed up with self-isolation during lockdown and don't want to risk any more of it. In a protest, a low bit-rate communications channel may be useful, not just for peer-to-peer messaging, but for transmitting instructions from organisers. In olden times, such instructions were broadcast by trumpeters or by pipers -- who quickly became targets for enemy archers or musketeers. BatNet gives better bandwidth than traditional acoustic battlefield signaling; it is partially jam resistant and also partially covert. If demo organisers wish to mask up and remain anonymous so they don't get targeted by snatch squads, some such communications are needed, and if the enemy controls the RF spectrum, then acoustics can provide an alternative.


\section{Conclusions}

\noindent In this paper, we introduced BatNet, an ultrasound communication mechanism between mobile devices that using the existing deployed hardware. 

\noindent We show that it is indeed possible to use ultrasound communication in standard smartphones for tactical communication. BatNet performs better than Google Nearby and is comparable to Lisnr. 

\noindent We analysed the possibility of using the BatNet mechanisms in covid contact tracing, and concluded that it did not add enough to the better-known known BLE mechanisms to be worth further development effort. However for tactical communications in the presence of a hostile state adversary -- such as by demonstrators whose mobile phone communications are jammed by the police -- it may offer significant benefits.

\section*{Acknowledgements}

\textit{Partially supported with funds from Bosch-Forschungsstiftung im Stifterverband.}

\bibliographystyle{unsrt}
\bibliography{references}

\begin{thebibliography}{1}

\bibitem{7961950}
D.~Arp, E.~Quiring, C.~Wressnegger, and K.~Rieck.
\newblock Privacy threats through ultrasonic side channels on mobile devices.
\newblock In {\em 2017 IEEE European Symposium on Security and Privacy (EuroS
  P)}, pages 35--47, April 2017.

\bibitem{Proakis}
John~G. Proakis and Masoud Salehi.
\newblock {\em Digital Communications}.
\newblock McGraw-Hill, 5th edition, 2008.

\bibitem{mcdaniel_mcdaniel}
Bill McDaniel.
\newblock An algorithm for error correcting cyclic redundancy checks.
\newblock {\em Dr. Dobb's}, June 2003.

\bibitem{8080245}
P.~{Getreuer}, C.~{Gnegy}, R.~F. {Lyon}, and R.~A. {Saurous}.
\newblock Ultrasonic communication using consumer hardware.
\newblock {\em IEEE Transactions on Multimedia}, 20(6):1277--1290, June 2018.

\bibitem{Lovett2020}
Tom Lovett, Mark Briers, Marcos Charalambides, Radka Jersakova, James Lomax,
  and Chris Holmes.
\newblock Inferring proximity from bluetooth low energy rssi with unscented
  {K}alman smoothers.
\newblock {\em Alan Turing Institute}, 2020.

\bibitem{Bounds2020}
Andy Bounds.
\newblock Outbreaks highlight disparities in {UK} test and trace regimes.
\newblock {\em Financial Times}, July 30 2020.

\bibitem{Anderson-contact2020}
Ross Anderson.
\newblock Contact tracing in the real world.
\newblock {\em Light Blue Touchpaper}, April 2020.

\end{thebibliography}

\end{document}